\documentclass[twocolumn,showpacs,pra]{revtex4}
\usepackage{amssymb}
\usepackage{graphicx}
\usepackage{dcolumn}
\usepackage{bm}
\usepackage{amsmath}
\usepackage{epsfig}

\begin{document}

\title{Laser Induced Selective Alignment of Water Spin Isomers}

\author{E. Gershnabel}
\author{I.Sh. Averbukh}
\affiliation{Department of Chemical Physics, The Weizmann Institute
of Science, Rehovot 76100, ISRAEL}
\begin{abstract}

We consider laser alignment of ortho and para spin isomers of water
molecules by using strong and short off-resonance laser pulses. A
single pulse is found to create a distinct transient alignment and
antialignment of the isomeric species. We suggest selective
alignment of one isomeric species (leaving the other species
randomly aligned) by a pair of two laser pulses.

\end{abstract}

\maketitle

\section{Introduction} \label{Introduction}

According to quantum mechanics, a molecule that contains two
identical atoms whose nuclei have nonzero spins can exist in the
form of spin isomers. In particular, a water molecule exists in one
of the two spin isomers, ortho or para with parallel or antiparallel
proton spins, respectively.  The ortho-para conversion is highly
improbable in the gas phase \cite{Bunker}. Water isomers separation
(or their ratio measurement) can be of great importance in medicine,
biology, chemistry,   magnetic resonance imaging, etc. Measurement
of the para to ortho ratio was used  to estimate the formation
temperature of molecules in space \cite{Dickens}. A method for
mixture enrichment in one of the water isomers was reported using
selective adsorption onto various surfaces \cite{Konyukhov},
although some of these results were hardly reproducible. Recently
Andreev $et$ $al$ \cite{Andreev} discussed an expected difference in
the response of the ortho and para forms of water to a non-uniform
dc electric field near solid surfaces, which may potentially lead to
isomers separation. We are currently exploring an alternative route
to ortho and para separation based on selective field-free alignment
and antialignment of water isomers by short nonresonant laser pulses
(for a review on field-free alignment see \cite{Stapelfeldt}). By
alignment we mean the angular localization of a molecular axis along
the laser polarization direction, while antialignment assumes
localization of the same molecular axis in the plane perpendicular
to the laser polarization. Although previous experimental studies on
dissociative fragmentation  by relatively long laser pulses have not
found alignment of water molecules \cite{Bhardwaj}, we show that a
sizable alignment can be achieved in ultra-short regime. The
selective alignment was recently experimentally demonstrated in a
mixture of ortho and para modifications of linear molecules
($^{15}N_2$ considered as an example) \cite{Sharly}. Here, for the
first time to our knowledge, we consider selective alignment for
bent tri-atomic molecules like water. The paper is organized as
follows. We  briefly discuss the classification of the states of the
ortho and para water molecules in Sec. \ref{WaterClassification}.
The interaction of the water molecules with the laser field is
considered in Sec. \ref{LaserMoleculeInteraction}. The results are
presented in Sec. \ref{Results} and  briefly summarized in Sec.
\ref{Conclusions}. In the Appendix, we give a detailed description
of the rotational wavefunctions of the ortho and para water isomers.

\section{Classification of the eigenstates of the Water Molecule} \label{WaterClassification}
The water molecule wavefunction is given by:

\begin{equation}
|\Psi\rangle=|\Psi_{rot}\rangle|\Psi_{spin}\rangle|\Psi_{vib}\rangle|\Psi_{ele}\rangle\
\ ,\label{EntireWavefunction}
\end{equation}
where $|\Psi_{rot}\rangle,|\Psi_{spin}\rangle,|\Psi_{vib}\rangle$
and $|\Psi_{ele}\rangle$ are rotational, nuclear spin, vibrational
and electronic wavefunctions, respectively. The water molecule
eigenstates are classified using the $C_{2v}(M)$ Molecular Symmetry
(MS) group \cite{Bunker}, i.e. using the irreducible representations
$A_1,A_2,B_1,B_2$. We provide the group operations as well as its
character table in the Appendix. According to the Pauli Exclusion
Principle, the total wavefunction should change its sign after
permutation of the hydrogen nuclei. Since we consider
$|\Psi_{vib}\rangle$ and $|\Psi_{ele}\rangle$ in the ground state
(which is symmetric with respect to the $C_{2v}(M)$ operations),
only $|\Psi_{rot}\rangle$ and $|\Psi_{spin}\rangle$ are effected by
the operations.
 The rotational part of
the water molecule Hamiltonian is presented by the rigid rotor
model:
\begin{equation}
\hat{H}=\frac{\hat{J}_a^2}{2I_a}+\frac{\hat{J}_b^2}{2I_b}+\frac{\hat{J}_c^2}{2I_c}\
\ ,\label{RotorHamiltonian}
\end{equation}
where $\hat{J}$ is the angular momentum operator, ($a,b,c$) are the
molecule's principal axes, and $I_a,I_b,I_c$ are the corresponding
moments of inertia ( $I_a<I_b<I_c$). Rotational states of $A_1$ and
$A_2$ symmetries do not change sign as a result of permutation and
belong to the para states (with anti-parallel spins), while the
states of $B_1$ and $B_2$ symmetries correspond to the ortho states.
Further information about the water molecule eigenstates
classification is given in the Appendix.

\section{ Interaction of the Water Molecules With  Laser Field} \label{LaserMoleculeInteraction}

\begin{figure}[htb]
\begin{center}
\includegraphics[width=8cm]{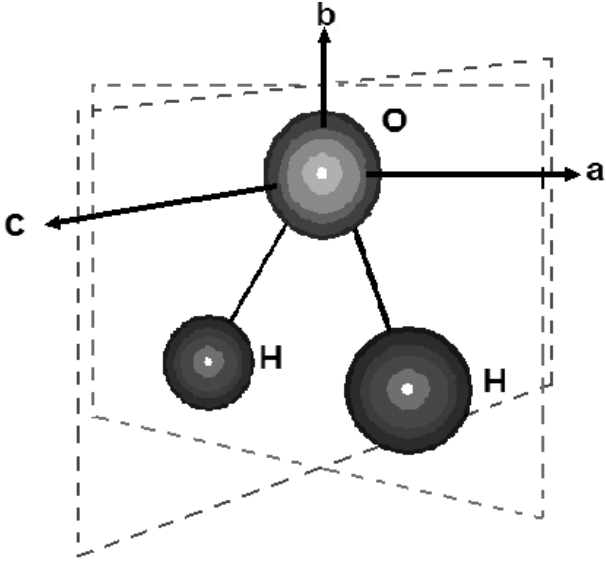}
\end{center}
\caption{The water molecule geometry. The axes are the principal
axes of the molecule, $a$,$b$ and $c$ ($a$ and $b$ are in the
molecular plane, and $c$ is perpendicular to it). The principal
moments of inertia are: $I_a=1.025\cdot10^{-47}kg\cdot m^2$,
$I_b=1.921\cdot10^{-47}kg\cdot m^2$ and
$I_c=2.946\cdot10^{-47}kg\cdot m^2$ (note that $I_c = I_a + I_b$
according to \cite{LandauAndLifshitz} )} \label{WaterMolecule}
\end{figure}

We consider the alignment process based on interaction with short
off-resonance laser pulses that induce molecular polarization,
interact with it, and excite rotational wavepackets. The interaction
of the water molecule (shown in Fig. \ref{WaterMolecule}) with the
laser pulse is given by \cite{Seideman}:
\begin{equation}
H_{int}=-\frac{1}{4}\sum_{\rho,\rho'}\varepsilon_{\rho}(t)\alpha_{\rho,\rho'}\varepsilon_{\rho'}(t)
 \ \ ,\label{Interaction}
\end{equation}
where the $\rho$ index denotes the space-fixed Cartesian
coordinates, $\varepsilon_{\rho}$ is the laser pulse envelope
component, and $\alpha$ is the polarizability tensor.

For linearly polarized fields oriented in the laboratory $Z$
direction, the interaction term is reduced to:

\begin{eqnarray}\label{InducedInteraction}
H_{ind}&=&-\frac{\varepsilon^2(t)}{4}[\alpha^{ab}\cos^2\theta+\alpha^{cb}\sin^2\theta\sin^2\chi]
\nonumber\\
&=&-\frac{\varepsilon^2(t)}{4}\{\frac{\alpha^{ab}+\alpha^{ac}}{3}D^2_{00}(\hat{R})
\nonumber\\
&-&\frac{\alpha^{cb}}{\sqrt{6}}[D^2_{02}(\hat{R})+D^2_{0-2}(\hat{R})]\}\
\ .
\end{eqnarray}

\begin{figure}[htb]
\begin{center}
\includegraphics[width=8cm]{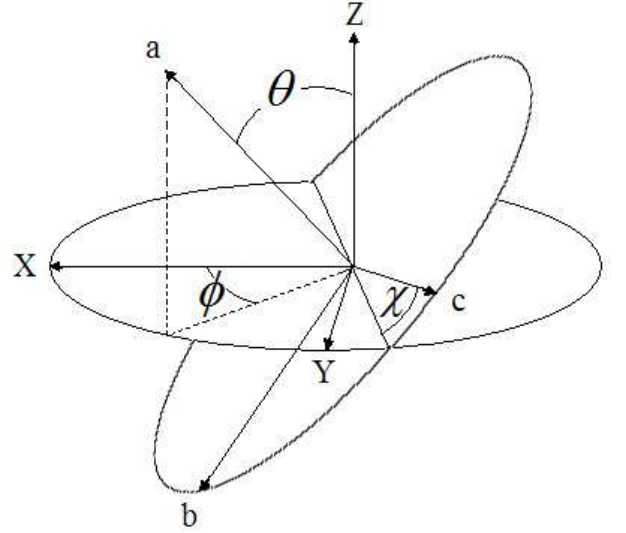}
\end{center}
\caption{The coordinate system. ($X,Y,Z$) are the space fixed
laboratory axes, ($a,b,c$) are the body fixed principal axes, and
($\theta,\phi,\chi$) are the Euler angles which define the body
orientation in space with respect to the space fixed coordinates}
\label{CoordinateSystem}
\end{figure}

The Euler angles $\theta$ and $\chi$, as well as the principal axes
of the molecule ($a,b,c$) appear in Fig. \ref{CoordinateSystem} that
 presents the coordinates used in this work. In Eq.
(\ref{InducedInteraction}),  $\alpha^{ab}=\alpha_{aa}-\alpha_{bb}$,
etc., $D_{mk}^J$ are the rotational matrices \cite{Zare}, and
$\hat{R}\equiv(\theta,\phi,\chi)$.

In the case of a symmetric top molecule (i.e. when
$\alpha_{xx}=\alpha_{yy}=\alpha_{\bot}$ and
$\alpha_{zz}=\alpha_{\|}$),   the Hamiltonian is reduced to the well
known form:
$H_{ind}=-\frac{1}{4}\varepsilon^2(t)\Delta\alpha\cos^2\theta$,
where $\Delta\alpha=\alpha_{\|}-\alpha_{\bot}$. In the case of
water, the polarizability tensor components are:
$\alpha_{aa}=1.528{\AA}^3$, $\alpha_{bb}=1.468{\AA}^3$ and
$\alpha_{cc}=1.415{\AA}^3$ \cite{Murphy}.

Given an initial rotational eigenstate of the asymmetric water
molecule, we are interested in the field-free evolution of the
rotational subsystem after an excitation by a laser pulse that is
short compared to the typical rotational periods of the water
molecule. Using the impulsive approximation,  we consider only the
contribution from the interaction term to the Schr\"{o}dinger
equation during the pulse:
\begin{equation}
\{\gamma_1D^2_{00}+\gamma_2[D^2_{02}+D^2_{0-2}]\}\Psi=i\hbar\frac{d\Psi}{dt}\
\ ,\label{Schrodinger}
\end{equation}
where $\gamma_1$ and $\gamma_2$ are found from Eq.
(\ref{InducedInteraction}).

Solving Eq. (\ref{Schrodinger}), we obtain the relation between the
states before and after the pulse:

\begin{eqnarray}\label{RelationBeforeAfter}
\Psi(t=0^+)=
\nonumber\\
\exp\left\{\frac{1}{i}\left\{\beta_1D^2_{00}+\beta_2[D^2_{02}+D^2_{0-2}]\right\}\right\}\Psi(t=0^-)
\ \ ,
\end{eqnarray}
where $\beta_{1,2}=(1/\hbar)\int\gamma_{1,2}dt$.

In order to find $\Psi(t=0^+)$, we introduce an artificial parameter
$\xi$ that will be assigned the value $\xi$=1 at the end of the
calculations:

\begin{eqnarray}\label{EqModified}
\Psi_{\xi}=
\nonumber\\
\exp\left\{\frac{1}{i}\left\{\beta_1D^2_{00}+\beta_2[D^2_{02}+D^2_{0-2}]\right\}\xi\right\}\Psi(t=0^-)\
\ .
\end{eqnarray}

We  express $\Psi_{\xi}$ in terms of the water molecule eigenstates:
\begin{equation}
\Psi_{\xi}=\sum_{J,\tau}c^+_{J,\tau}|J,\tau,m\rangle\ \ ,
\label{Expand}
\end{equation}
where $J$ is the angular momentum quantum number, and $m$ is the
quantum number of the angular momentum projected on the laboratory
$Z$ axis. For the chosen laser field polarization, $m$ is a
conserved quantum number. This allows us to consider the excitation
of the states with a different initial $m$-value separately. Index
$\tau$ (which is assigned the values $-J,...,+J$) numerates various
$2J+1$ eigenstates arranged from the lowest energy ($\tau=-J$) to
the highest energy ($\tau=+J$).

Then we take a derivative of both sides of Eq. (\ref{EqModified})
with respect to $\xi$, apply the bra $\langle J',\tau ',m|$ from the
left, and  obtain a set of differential equations:

\begin{eqnarray}\label{DifferentialEquations}
\dot{c}_{J'\tau
'}^+=\frac{1}{i}\sum_{J,\tau}c_{J,\tau}^+\{\beta_1\langle J'\tau '
m|D^2_{00}|J \tau m\rangle
\nonumber\\
+ \beta_2\langle J' \tau ' m|D^2_{02}|J \tau m\rangle
+\beta_2\langle J' \tau ' m|D^2_{0-2}|J \tau m\rangle\}\ \ ,
\end{eqnarray}
where $\dot{c}\equiv dc/d\xi$.

Since $\Psi_{\xi=0}=\Psi(t=0^-)$ and $\Psi_{\xi=1}=\Psi(t=0^+)$ (Eq.
(\ref{EqModified})), we solve numerically this set of equations from
$\xi=0$ to $\xi=1$, and find $\Psi(t=0^+)$.

The post-pulse field-free propagation of the created rotational wave
packet is calculated as
\begin{equation}
\Psi(t)=\sum_{J,\tau}c^+_{J,\tau}\exp\left\{{-i\frac{E^{J
\tau}}{\hbar}t}\right\}|J,\tau,m\rangle \ \ ,\label{propagate}
\end{equation}
where $E^{J\tau}$ is the eigenvalue corresponding to the eigenstate
$|J,\tau,m\rangle$. We calculate the matrix elements in Eq.
(\ref{DifferentialEquations}) by expressing the water molecule
eigenstates using the basis of the symmetric rotor: $|J, k,m\rangle$
($J$ and $m$ are defined as before, and $k$ is the quantum number of
the angular momentum projected on the rotor symmetry axis), and
using the following relations:

\begin{eqnarray}\label{UsefulRelation}
\langle J k m|D^2_{0s}|J' k' m\rangle=(-1)^{k'+m}\times
\nonumber\\
\sqrt{(2J+1)(2J'+1)}
\begin{pmatrix}   j & 2 & j'\\   m & 0 & -m
\end{pmatrix}\begin{pmatrix}   j & 2 & j'\\   k & s & -k'
\end{pmatrix}\ \ .
\end{eqnarray}

Here $s=0,\pm2$, and $\begin{pmatrix}   j_1 & j_2 & j_3\\   m_1 &
m_2 & m_3
\end{pmatrix}$ is the $3-j$ symbol, related to the Clebsch-Gordan coefficient by:

\begin{eqnarray}\label{Clebsch}
\begin{pmatrix}   j_1 & j_2 & j_3\\   m_1 & m_2 & m_3
\end{pmatrix}=(-1)^{j_1-j_2-m_3}(2j_3+1)^{-1/2}\times
\nonumber\\
\langle j_1 m_1 j_2 m_2|j_3 -m_3\rangle\ \ .
\end{eqnarray}
The non-zero matrix elements from  Eq. (\ref{UsefulRelation})
satisfy the following selection rules: $|J-2|\leq J' \leq J+2$ and
$k' =k,k\pm2$.

In order to evaluate the values of $\beta_{1,2}$, we consider the
following  pulse shape: $\varepsilon^2(t)=\varepsilon_0^2
\exp\left\{\frac{-t^2}{2\sigma^2}\right\}$. Considering the maximal
pulse intensity to be $I=3\cdot 10^{13} W/cm^2$ and duration
$\sigma=20fs$ we evaluate $\varepsilon_0$ as $\varepsilon_0=1.5\cdot
10^{10} V/m$ and obtain $\beta_1=-0.174$ and $\beta_2=-0.065$.

In order to calculate the alignment factor $\langle
\cos^2\theta\rangle$ at temperature $T$, one has to perform thermal
averaging:
\begin{equation}
\langle \cos^2 \theta \rangle_T(t) =\sum_{J,\tau,m}\frac{\exp\{-E^{J
\tau}/k_BT\}}{Q_{rot}}\langle \cos^2 \theta\rangle_{J \tau m}(t)\ \
.\label{ThermalAvg}
\end{equation}

Here $Q_{rot}$ is the rotational partition function, $k_B$ is the
Bolzmann's constant, $\langle \rangle_T$ denotes quantum thermal
average over the molecular ensemble, and the indexes $J,\tau,m$
denote the initial rotational state $|J,\tau,m\rangle$ of a molecule
in the ensemble. The quantum average is calculated using:
\begin{equation}
\langle \cos^2 \theta\rangle_{J \tau m}(t)=\langle \Psi(t)
|\cos^2\theta|\Psi(t)\rangle\ \ ,\label{cosSquare}
\end{equation}
where $\Psi(t)$ is found by solving Eqs. (\ref{Expand}),
(\ref{DifferentialEquations}) and (\ref{propagate}) with the initial
state $\Psi(t=0^-)=|J,\tau,m\rangle$. The right hand side in Eq.
(\ref{cosSquare}) can be easily calculated using the relation:
\begin{equation}
\cos^2\theta=\frac{2D^2_{00}(\hat{R})+1}{3}\ \ .
\end{equation}

It is natural to consider the alignment of the para and ortho water
molecules separately, since the laser pulse does not mix the para
and ortho rotational states, i.e. if one considers an initial para
rotational state and apply the pulse, only para rotational states
are excited. The same holds for ortho rotational states. This is
because the interaction term, $H_{ind}$, in Eq.
(\ref{InducedInteraction}) includes:
\begin{eqnarray}\label{RotationalMatrices}
D^2_{00}(\hat{R})&\propto& \langle\hat{R}|J=2,k=0,m=0\rangle
\nonumber\\
D^2_{02}(\hat{R})+D^2_{0-2}(\hat{R}) &\propto& \langle
\hat{R}|2,2,0\rangle+\langle\hat{R}|2,-2,0\rangle,
\end{eqnarray}
which (according to Eq. (\ref{GroupOperations}) in the Appendix)
have $A_1$ symmetry. In order to have a transition from a para to
ortho state, the following matrix element should not necessarily
vanish: $\langle Para|H_{ind}|Ortho\rangle$. Since the interaction
term has a symmetry $\Gamma_{ind}=A_1$, the para state has
$\Gamma_{para}=A_1$ or $A_2$ symmetries, and the ortho has
$\Gamma_{ortho}=B_1$ or $B_2$, direct product of the symmetries
$\Gamma_{para}\times \Gamma_{ind}\times\Gamma_{ortho}$ never
contains $A_1$ and therefore necessarily $\langle
Para|V_{ind}|Ortho\rangle=0$ (forbidden transition). Thus, the
transitions between para and ortho states are forbidden and the sum
in Eq. (\ref{ThermalAvg}) includes either initial para states or
initial ortho states separately.

\section{Results} \label{Results}

\begin{figure}[htb]
\begin{center}
\includegraphics[width=8cm]{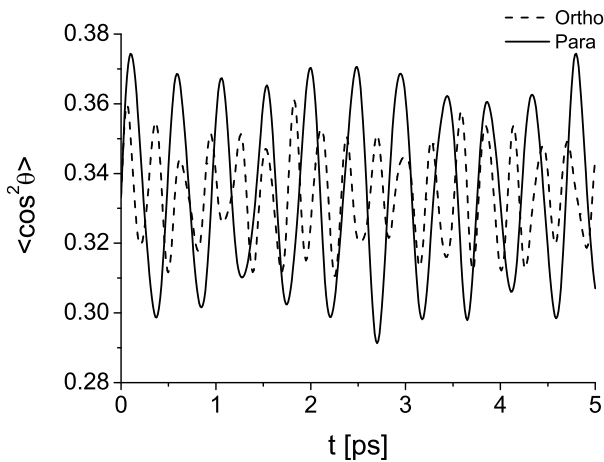}
\end{center}
\caption{The alignment factor after the laser pulse at  temperature
of $20K$. Different dynamics of the ortho and para isomers is
revealed during free propagation of the rotational wavepackets.}
\label{WaterAlignmentGraph}
\end{figure}

Applying a linearly polarized $3\cdot 10^{13}W/cm^2$, $20fs$ laser
pulse, we expect the molecular highest polarizability axis (i.e. the
$a$ axis in Fig. \ref{WaterMolecule}) to be aligned along the laser
field polarization shortly after the pulse. Since the pulse is short
compared to the  typical rotational periods of the water molecule,
it is considered as a delta-pulse.  The calculated alignment factor
$\langle\cos^2\theta\rangle_T$ (Eq. (\ref{ThermalAvg})) is plotted
in Fig. \ref{WaterAlignmentGraph} as a function of time at
temperature of  $20K$.

\begin{figure}[htb]
\begin{center}
\includegraphics[width=8cm]{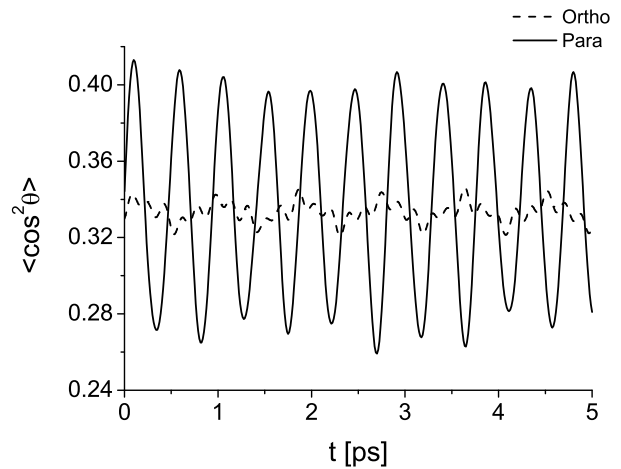}
\end{center}
\caption{Alignment factor after the second pulse versus time at
$20K$. While the transient alignment of the para molecules is
increased, the ortho alignment is reduced, and the ortho molecules
become almost isotropically aligned.}
\label{WaterSingleAlignmentGraph}
\end{figure}

It is seen that the ortho and para rotational wavepackets evolve
differently after the application of the pulse, and transient
moments of simultaneous alignment and anti-alignment for the two
species can be found. For instance, at $t=2\ ps$, the para molecules
are aligned while the ortho molecules become anti-aligned.
Naturally, the rotational energies of both ortho and para molecules
are increased as the result of the pulse. Further selective control
of the isomers can be achieved by applying an additional pulse (of
the same intensity and duration, for simplicity) at $t=1.9\ ps$,
when the ortho and para molecules are on the way of becoming
anti-aligned and aligned, respectively. At this time, the alignment
factors of both isomers are close to each other, and are a bit
larger than $1/3$. The dynamics   of the alignment factors after the
second pulse is shown in Fig. \ref{WaterSingleAlignmentGraph}
(again, for temperature of $20K$).

It depicts a selective alignment (and anti-alignment) of the para
molecules, while leaving the ortho modifications more or less
isotropically distributed. Moreover, the total rotational energy of
the ortho and para molecules is affected in a different way by the
second pulse. While the rotational energy of the para molecules is
further increased, the energy of the ortho molecules is reduced as
the result of the action of the second pulse.

\section{Conclusions} \label{Conclusions}

We considered interaction of the water molecules with strong and
short off-resonant laser pulses and analyzed dynamics of rotational
wavepackets of ortho and para spin isomers of water. We showed that
these two forms of the water molecules experience distinct series of
transient alignment and anti-alignment events, which leads to the
angular separation of an initially isotropic mixture of nuclear spin
modifications.  Moreover, we showed that by applying a second,
properly delayed laser pulse it is possible to suppress the
rotational energy of one isomeric species, and leave it in a more or
less isotropic state. On contrast, the other species will experience
an enhanced series of transient alignments and anti-alignments.
 Such a selective alignment establishes grounds for the
actual separation of ortho and para water isomers via interaction
with additional spatially inhomogeneous static or time-dependent
fields. This is a subject of our ongoing research.

We acknowledge the support of the Israel Science Foundation. This
research is made possible in part by the historic generosity of the
Harold Perlman Family. IA is an incumbent of the Patricia Elman
Bildner Professorial Chair.

\section{Appendix- Rotational Wavefunctions Classification} \label{AppendixA}

The eigenstates of the water molecule are classified using the
$C_{2v}(M)$ Molecular Symmetry (MS) group \cite{Bunker}. Considering
the water molecule eigenstate (Eq. (\ref{EntireWavefunction})), we
define the nuclei rotational and vibrational, as well as electronic
degrees of freedom as a rovibronic eigenstate. We denote the
rovibronic eigenstate symmetry as $\Gamma_{rve}$ (within the MS
group), and the nuclear spin state symmetry as $\Gamma_{ns}$ (within
the MS group). The water nuclear spin states are divided into either
$S=0$ or $1$ nuclear spin states, denoted as para or ortho,
respectively. The Pauli principle imposes symmetry limitations on
the valid basis functions for expressing the internal molecular
wavefunction, $\Phi_{int}$ (the entire molecular wavefunction,
without the translational degrees of freedom). A rovibronic state
having  $\Gamma_{rve}$ symmetry can be only combined with a nuclear
spin state having  $\Gamma_{ns}$ symmetry if the product of the two
symmetries is an allowed symmetry for $\Phi_{int}$.  In this
Appendix we  review the symmetries of the rotational eigenstates,
and identify each rotational eigenstate as para or ortho.
\\
\\
We  start by examination of the symmetry of the internal
wavefunction, $\Phi_{int}$. The character table of the MS group is
given in Table \ref{character}. The group operations are: $E$,
$(12)$, $E^*$ and $(12)^*$, where $E$ is the identity operator,
$(12)$ is the permutation of the Hydrogen nuclei operator, $E^*$ is
the inversion operator (both for the nuclei and electron
coordinates) and $(12)^*$ is the combination $(12)E^*$. In Table
\ref{character} only the symmetries $B_2$ (positive parity with
respect to inversion) and $B_1$ (negative parity with respect to
inversion) change sign under the operation $(12)$ (i.e. the
permutation of the two Hydrogen nuclei, which are fermions).
Therefore, they are the only allowed symmetries of the internal
wavefunction.
\\
\\
Next, we  classify the nuclear spin wavefunction. There are two
possible spin functions for  $s=1/2$ proton spin:
$\alpha=|\frac{1}{2},\frac{1}{2}\rangle$ and
$\beta=|\frac{1}{2},-\frac{1}{2}\rangle$, where $\alpha$ and $\beta$
correspond to spin up and spin down states, respectively. There are
four possible spin configurations:

\begin{eqnarray} \label{spinors}
m=1:\alpha\alpha=\Phi_{ns}^{(1)} \nonumber\\
m=0:\alpha\beta=\Phi_{ns}^{(2)};\beta\alpha=\Phi_{ns}^{(3)}\nonumber\\
m=-1: \beta\beta=\Phi_{ns}^{(4)}\ \ .
\end{eqnarray}

One should bear in mind that the spin functions are invariant to
$E^*$, and also for them $(12)^*=(12)$. $\Phi_{ns}^{(1)}$ and
$\Phi_{ns}^{(4)}$ are invariant to the group operations and have
symmetry $A_1$ (see Table \ref{character}). The characters of
$\Phi_{ns}^{(2)}$ and $\Phi_{ns}^{(3)}$ (which transform together)
are $2,0,2,0$ (for $E$,$(12)$,$E^*$,$(12)^*$, respectively), i.e.
$\Phi_{ns}^{(2)}$ and $\Phi_{ns}^{(3)}$ generate the representation
$A_1\oplus B_2$. The four nuclear spin wavefunctions therefore
generate the representation $\Gamma_{ns}=3A_1\oplus B_2$.

\begin{table}[ht]
\centering
\begin{tabular}{c c c c c}
\hline\hline $C_{2V}(M)$ & $E$ & $(12)$ & $E^*$ & $(12)^*$ \\
$Rotation$ & $R^0$ & $R_b^{\pi}$ & $R_c^{\pi}$ & $R_a^{\pi}$\\
[0.5ex] \hline
$A_1$ & $1$ & $1$ & $1$ & $1$\\
$A_2$ & $1$ & $1$ & $-1$ & $-1$\\
$B_1$ & $1$ & $-1$ & $-1$ & $1$\\
$B_2$ & $1$ & $-1$ & $1$ & $-1$\\
 [1ex] \hline
\end{tabular}
\caption{The character table of the $C_{2v}(M)$ MS group. $E$ is the
identity operator, $(12)$ is the permutation of the Hydrogen nuclei
operator, $E^*$ is the inversion operator (both for the nuclei and
electron coordinates) and $(12)^*$ is the combination $(12)E^*$. The
effect of the MS group on the rotational eigenstates is equivalent
to the effect of the operations of the molecular rotation group
$D_2$ (i.e. $R^0,R_b^{\pi},R_c^{\pi},R_a^{\pi}$) on the rotational
eigenstates.} \label{character}
\end{table}

As the next step, we  classify each rotational eigenstate. It turns
out that the effect of the MS group on the rotational eigenstates is
equivalent to the effect of the operations of the molecular rotation
group $D_2$ (i.e. $R^0,R_b^{\pi},R_c^{\pi},R_a^{\pi}$) on the
rotational eigenstates (see Table \ref{character}). $E$ or $R^0$ is
the identity operation. $(12)$ (permutation of the two Hydrogen
nuclei) is a rotation about $b$ by $\pi$ (see Fig.
\ref{WaterMolecule}), which we denote as $R_b^{\pi}$. $E^*$ is
rotation about $c$ by $\pi$, i.e. $R_c^{\pi}$. Finally $(12)^*$ is a
rotation about $a$ by $\pi$, i.e. $R_a^{\pi}$. The operations are
given by \cite{Bunker}:

\begin{eqnarray}\label{GroupOperations}
R_a^{\beta}|J,k,m\rangle &=& e^{ik\beta}|J,k,m\rangle
\nonumber\\
R_{\alpha}^{\pi}|J,k,m\rangle &=& (-1)^Je^{-2ik\alpha}|J,-k,m\rangle
\ \ ,
\end{eqnarray}
where $\alpha=0,\pi/2$ for $b$ and $c$, respectively. Using Eq.
(\ref{GroupOperations}) for each eigenstate, one can find the
characters of the group operators operating on the eigenstate,
compare them to Table \ref{character}, and identify the symmetry of
the eigenstate. The results are given in Table \ref{ParaOrtho}
(presented here only for $J=0,1,2$ and $3$).

We assume the vibrational and electronic subsystems to be in the
ground state (that has $A_1$ symmetry). The overall rovibronic
symmetry is  therefore determined by the rotational symmetry.

One may form a valid basis wavefunction for expressing $\Phi_{int}$
by combining a rovibronic state having a $\Gamma_{rve}$ symmetry and
a nuclear spin state having a $\Gamma_{ns}$ symmetry, in such a way
that the product of these two  contains $\Gamma_{int}$
($\Gamma_{rve}\otimes\Gamma_{ns}\supset \Gamma_{int}$, where
$\Gamma_{int}$ is an allowed symmetry for $\Phi_{int}$). A detailed
list of rovibronic symmetries and their corresponding spin
symmetries is given in Table \ref{StatisticalWeight}.

\begin{table}[ht]
\centering
\begin{tabular}{c c c}
\hline\hline $\Gamma_{rve}$ & $\Gamma_{ns}$ & $\Gamma_{int}$\\
[0.5ex] \hline
$ A_1 $ & $ B_2;- $ & $ B_2;B_1 $\\
$ A_2 $ & $ -;B_2 $ & $ B_2;B_1 $\\
$ B_1 $ & $ -;3A_1 $ & $ B_2;B_1 $\\
$ B_2 $ & $ 3A_1;- $ & $ B_2;B_1 $\\
 [1ex] \hline
\end{tabular}
\caption{The symmetries of the rovibronic eigenstates appear in the
left column. The right column displays the possible symmetries of
the internal eigenstates. The middle column presents the
corresponding nuclear spin symmetries such that
$\Gamma_{rve}\otimes\Gamma_{ns}\supset \Gamma_{int}$. $\Gamma_{ns}$
can be either $B_2$ or $3A_1$.  The $'-'$ sign indicates  that no
appropriate $\Gamma_{ns}$ exists for the combination of symmetries
in the first and the third columns.} \label{StatisticalWeight}
\end{table}

From Table \ref{StatisticalWeight} one can relate the $A_1$ and
$A_2$ rovibrational wavefunctions with $S=0$ nuclei spin, i.e. para
water (the antisymmetric $S=0$ wavefunction has a $B_2$ symmetry).
Similarly, the $B_1$ and $B_2$ rovibrational wavefunctions are
related with the $S=1$ nuclei spin, i.e. ortho water (the symmetric
$S=1$ wavefunctions generate the representation $3A_1$). The
statistical weight is therefore clear from the Table: there are
three times more ortho species than para species ($75\%$ of the
water is ortho and $25\%$ is para).

Finally, Table \ref{ParaOrtho}  identifies  the eigenstates of water
molecule as para or ortho for various values of $J$ and $\tau$ .

\begin{table}[ht]
\centering
\begin{tabular}{c c c c}
\hline\hline $Para(P)/Ortho(O)$ & $Symmetry$ & $J$ & $\tau$\\
[0.5ex] \hline
$P$ & $ A_1 $ & $0$ & $0$\\
$O$ & $ B_1 $ & $1$ & $-1$\\
$P$ & $ A_2 $ & $1$ & $0$\\
$O$ & $ B_2 $ & $1$ & $1$\\
$P$ & $ A_1 $ & $2$ & $-2$\\
$O$ & $ B_2 $ & $2$ & $-1$\\
$P$ & $ A_2 $ & $2$ & $0$\\
$O$ & $ B_1 $ & $2$ & $1$\\
$P$ & $ A_1 $ & $2$ & $2$\\
$O$ & $ B_1 $ & $3$ & $-3$\\
$P$ & $ A_2 $ & $3$ & $-2$\\
$O$ & $ B_2 $ & $3$ & $-1$\\
$P$ & $ A_1 $ & $3$ & $0$\\
$O$ & $ B_1 $ & $3$ & $1$\\
$P$ & $ A_2 $ & $3$ & $2$\\
$O$ & $ B_2 $ & $3$ & $3$\\
 [1ex] \hline
\end{tabular}
\caption{Para and ortho wavefunctions (here presented only till
$J=3$) identification. The water asymmetric rotor eigenstates are
$|J,\tau,m\rangle$, where $J$ is the angular momentum quantum number
($J=0,1,2,...$), and $m$ is the quantum number of the angular
momentum projected on the laboratory $Z$ axis (since there exists no
preferred direction in space, we choose here $m=0$ without loss of
generality). Index $\tau$ numerates  various $2J+1$ eigenstates
arranged from lowest energy ($\tau=-J$) to highest energy
($\tau=+J$). The Table displays also the symmetry of the rotational
eigenstates.} \label{ParaOrtho}
\end{table}

\bibliographystyle{phaip}

\end{document}